\documentclass[aip, pop, amsfonts, amssymb, amsmath ,preprint,subscriptadress]{revtex4-1}
\usepackage{amssymb}		 
\usepackage{amsmath, amsthm, amssymb,amsfonts}
\usepackage[english]{babel}
\usepackage{color, graphicx}
\begin{document}

\title{{Generalised  relativistic Ohm's laws, extended gauge transformations  \\  and magnetic linking}}
\author{{F. Pegoraro}\\
 Dipartimento di Fisica, Universit\`a di Pisa, Italy}

\begin{abstract}
Generalisations of the  relativistic ideal Ohm's law are presented that include specific  dynamical  features  of the current carrying particles in a plasma.      Cases of  interest for space and laboratory plasmas are identified where these generalisations allow for  the definition of generalised electromagnetic fields that transform under a Lorentz boost in the same way as the real electromagnetic fields  and that obey the same set of homogeneous Maxwell's equations.
    
\end{abstract}

\maketitle

\section{Introduction}\label{intr}

In the fluid description of a plasma 
the  momentum equation of the lighter particle species, generally the electrons, plays a fundamental role in determining the  properties of the  large spatial scale,  low frequency dynamics.  This is particularly evident in the case of the single fluid MHD description where the so called ideal Ohm's law is essentially the momentum equation of massless, cold  electrons. More precisely  it expresses the vanishing of the electromagnetic force  on the electron fluid  under the additional assumption that the electron and the ion fluid velocities can be assumed, within this equation, to coincide.  In addition it is assumed  that there is no electron fluid momentum transmitted through collisions  to different species (resistivity) or to the electromagnetic fields through high frequency incoherent radiation (radiation friction).

 It is a fundamental feature of   MHD that, if the ideal Ohm's law applies,  the plasma dynamics is constrained by an infinite set of topological invariants, 
 such as the one arising from the conservation of magnetic helicity. These constraints express  the invariance  in time of {\it magnetic connections}, i.e., the property that if the ideal Ohm's law holds and the plasma velocity field remains smooth, two  fluid elements that  at $t=0$ are linked by a magnetic field line remain linked by a magnetic field line at any successive time \cite{Newc}. This property goes under the  abbreviated  but suggestive statement  that the magnetic field is frozen in the plasma.
\\
It is also well known that a number of physical effects leads to violations of the ideal Ohm's and that these effects are generally related to the appearance of small spatial and/or temporal scales due  e.g.,  to  the nonlinearity of the plasma dynamics.  When these violations occur only locally,  the magnetic field lines in the plasma undergo the well known process of magnetic reconnection. In this process the identity of fields lines is lost only inside the reconnection region  but the linking between different fluid elements is changed globally, causing a rearrangement of the global magnetic field topology.

There are different ways in which the ideal Ohm's law can be violated: they can be roughly grouped into three different classes.  In a first  class the violation amounts to a change of  the fluid with respect to which the magnetic field is frozen, as is the case of a two-fluid plasma description where the restriction that the ion fluid and the electron fluid move with the same  velocity  is relaxed,  or in the so called Electron Magnetohydrodynamics \cite{Kurch,EMHD}  where ions are taken to be immobile. In this class, which includes the so called Hall-MHD \cite{hallMHD}, the magnetic field remains frozen  with respect to the electron fluid. 
As a consequence, the topology of the magnetic field  ${\vec B}$ is preserved although the ion fluid is allowed to slip with respect to the magnetic field,

A second class involves a change in the fields that define the linking. This is the case when the  assumption of massless electrons is relaxed and thus work must be performed in order to accelerate them. In this case is was shown  (see  Ref.\onlinecite{Yan})  that, if the electron fluid is assumed to be cold,   a generalised magnetic field ${\vec B}_e \equiv \nabla \times ({\vec A} - e {\vec u}_e /m_e)$ is frozen in the electron fluid and  a generalised ideal Ohm's law can be written in the form
\begin{equation}\label{int1} 
{\vec E}_e  + {\vec u}_e \times {\vec B}_e/c = 0, \end{equation}
where ${\vec E}_e = - \nabla(\phi -  e| {\vec u}_e|^2 / (2 m_e)  - \partial_t ({\vec A} - e {\vec u}_e /m_e)/c$.
In addition the fields ${\vec E}_e $  and ${\vec B}_e $ satisfy the homogeneous Maxwell's equation $\nabla\cdot  {\vec B}_e = 0$ and  $\nabla \times {\vec E}_e = - \partial_t  {\vec B}_e /c$.
In this case the topology of the magnetic field  ${\vec B}$ is  not preserved but  the topology of the  generalised magnetic field  ${\vec B}_e$ is, i.e.,  ${B}_e$-connections  are preserved by the electron  dynamics.   In this case, see Ref.\onlinecite{de} and references therein,  magnetic reconnection can only proceed if large gradients of the electron fluid velocity  ${\vec u}_e$, or somewhat equivalently of the plasma current density, are produced.
A  similar result  applies  if we relax the  condition of cold electrons and introduce in Ohm's law  the gradient of an isotropic pressure that is a function of the plasma density only.  In the non relativistic case this can be performed  by  adding the   contribution arising from the gradient of the pressure  to the gradient of the electrostatic potential $\phi$.   In this case, if for example the electron inertia is neglected, the magnetic field ${\vec B}$ remains frozen in the plasma MHD flow.

The third class involves phenomena that  are the consequence of a momentum transfer   to the other particle species either through collisions or higher frequency collective phenomena, i.e., through the effect of 
collisional or anomalous resistivity. Electron momentum can also be lost  through high frequency incoherent radiation (the so called radiation reaction force or radiation friction),  or  spatially redistributed  between different electron fluid elements by electron viscosity. Additional violations in this class arise from  electron kinetic effects that are not accounted for within a standard  fluid description, such as Landau damping or an anisotropic and in particular  non-gyrotropic pressure tensor.   \, Contrary to the two previous cases, for this  class  it is not normally  possible to define a generalised  magnetic field that remains frozen in a fluid plasma component, however selected.
\bigskip

An important feature of the ideal Ohm's law is that it is in no sense restricted to a non relativistic plasma regime, as it can be written (unmodified) in the fully covariant form
\begin{equation}\label{int2}
F_{\mu\nu}u_\nu = 0,
\end{equation}
where $F_{\mu\nu}$ is the electromagnetic (e.m.)  field tensor, $u_\mu$ is a normalised timelike 4-vector ($u_\mu u_\mu = -1$) which we interpret as the fluid velocity  4-vector field of the plasma species with respect to which the magnetic field is frozen\cite{fn1}.
While the ideal Ohm's law is fully covariant, its  interpretation in terms of the conservation of magnetic connections is not.  In fact 
the meaning of magnetic connection and of magnetic topology is not  clear in a relativistic context  because of two related reasons:  first, the distinction between electric and magnetic
fields is frame dependent and second the very concept of field lines, which are defined in coordinate space at a given time, is frame dependent due to the violation of simultaneity in different reference frames of events at different spatial locations.  This feature was  addressed in Ref.\onlinecite{FPEPL} where it was shown that the  covariant formulation of magnetic connections  can be restored by means of a Òtime resettingÓ projection along the trajectories of the plasma elements. This projection is consistent with the ideal Ohm's law and   compensates  for the loss of simultaneity in different reference frames between spatially separated events.  It was then shown (see Ref.\onlinecite{Arcetri}) that  a frame independent definition of magnetic topology  can be recovered  by referring  to 2D-hypersurfaces in 4D Minkowski space instead of 1D curves in 3D space  at fixed time. These hypersurfaces are defined by the two linearly independent 4-vector fields\cite{fn2} whose contraction with the e.m.  field tensor $F_{\mu\nu}$ is identically zero, while the corresponding homogeneous Maxwell equations  $\partial_\mu  {\cal F}_{\mu\nu} = 0$, with $ {\cal F}_{\mu\nu} \equiv  \varepsilon_{\mu\nu\lambda\sigma} { F}_{\lambda\sigma}/2 $ the dual of the e.m.  tensor ${ F}_{\mu\nu} $, play the role of a Frobenius  involution condition for the existence of the foliation  of Minkowski space defined by  these hypersurfaces.   The covariant definition of these hypersurfaces  makes it possible to define magnetic connection-lines covariantly by taking  cuts at the same  coordinate time in each reference frame.
\bigskip

In the present article we address the relativistic covariant formulation of a  non-ideal Ohm's law  (Sec.\ref{ROL}) and  look   for the conditions that are required in order to define a covariant form of generalised connections. In this context we note   the analysis  recently presented in Ref.\onlinecite{AsCo} where generalized magnetic connections  are derived for a set of relativistic non-ideal MHD equations that include thermal-inertial, thermal-electromotive, Hall and current-inertia effects. \\ We show that the  conditions  required in order to define a covariant form of generalised connections can be satisfied automatically by introducing a {\it generalised gauge transformation} of the 4-vector potential  $A_\mu$ defined by a {\it gauge field} $s_\mu$ that must satisfy a compatibility condition involving the 4-velocity $u_\mu$. \, We refer in particular to the case where inertial  and thermal electron effects are considered  (Sec.\ref{RIOL}).  The results obtained in this Section agree with  the  analysis in  Ref.\onlinecite{AsCo}  when  the difference between the  adopted plasma descriptions   is taken into account: generalized relativistic  MHD equations in   Ref.\onlinecite{AsCo},  relativistic electron fluid equations coupled to the homogeneous Maxwell's  equations in the present article.
\\
An interesting extension to a fluid of relativistic spherical tops is given in Sec.\ref{spin}  while two dissipative cases, radiation friction and collisional resistivity are discussed in Secs.\ref{RRc},\ref{CL} respectively. The definition of generalised helicity in given in Sec.\ref{hel} while the relevance of the present analysis to the development of magnetic reconnection is briefly discussed in the Conclusions. 
\bigskip

Before proceeding we recall that  the tensor contractions $F_{\mu\nu} F_{\mu\nu}$ and $F_{\mu\nu} {\cal F}_{\mu\nu}$ are Lorentz invariants proportional to $|{\vec E}|^2 - |{\vec B}|^2$  and to ${\vec E} \cdot {\vec B} $ respectively  and that  $F_{\mu\nu} {\cal F}_{\mu\nu}$ vanishes if  an  equation of the form of   Eq.(\ref{int2}) holds where, in general,  
the 4-vector field that is annihilated by the e.m. field tensor $F_{\mu\nu} $ need not be timelike.

\section{Relativistic Ohm's law}\label{ROL}

We write the relativistic Ohm's law  in formal terms as
\begin{equation}\label{1}
F_{\mu\nu}u_\nu = R_\mu,
\end{equation}
with  
$R_\mu $  a 4-vector field 
such that 
\begin{equation}\label{1a}
u_\mu R_\mu = 0.
\end{equation}
The  4-vector $R_\mu $  is taken to include any non ideal effect not included in Eq.(\ref{int2}).  Note however that  if  $R_\mu  $ can be put in the form $R_\mu = - F_{\mu\nu}v_\nu $ such that 
if $u_\mu  + v_\mu \not= 0$ is still a timelike 4-vector field, this  violation of the ideal Ohm's law  can in principle  be  removed by  a different choice of the ``reference'' 4-velocity\cite{fn3}.   For this to occur the Lorentz  invariant $F_{\mu\nu} {\cal F}_{\mu\nu}$  must vanish. This case will not be considered in the rest of this article 
as we will  require that the violation of the ideal Ohm's law makes $F_{\mu\nu} {\cal F}_{\mu\nu} \not= 0$  at least locally.

Using the standard  decomposition \cite{Li,Ani,D'A}  of the field tensor 
\begin{equation}\label{2} F_{\mu\nu}  =  {\varepsilon_{\mu\nu\lambda\sigma} b_{\lambda}u_{\sigma}}\,  + \,  {[u_{\mu}e_{\nu} - u_{\nu}e_{\mu}]} \, , \end{equation}
where \, {$b_\mu$}  is the {\it 4-vector magnetic  field} and {$e_\mu$}  is the {\it 4-vector electric field}, 
with  $u_\mu e_\mu =0$ and $u_\mu b_\mu =0$, we find 
\begin{equation}\label{2a}
R_\mu =  e_\mu.
\end{equation}
The  4-vectors $e_\mu$ and $b_\mu$ in Eq.(\ref{2}) are related to the standard electric and magnetic fields ${\vec E}$ and ${\vec B}$ in 3D space by
\begin{equation}\label{2b}
b_\mu = \gamma({\vec B} \cdot  {\vec v}\,  , \, {\vec B}  + {\vec E} \times {\vec v}  ), 
\end{equation}
and 
\begin{equation}\label{2c}
e_\mu = \gamma({\vec E} \cdot  {\vec v}\,  , \, {\vec E} + {\vec v} \times {\vec B}  ) ,
\end{equation}
with $e_\mu b_\mu =  {\vec E}  \cdot {\vec B}$, \, $\gamma$ is the relativistic Lorentz factor and we have used $ u_\mu = \gamma (1,  {\vec v})$.\\
The representation given in Eq.(\ref{2}) is physically convenient as it allows us to separate  covariantly  the magnetic and the  electric  parts of the e.m. field tensor relative to a given  plasma component moving  with  4-velocity $u_\mu$.  In the  local rest frame of this plasma component   the time components of $e_\mu$ and of $b_\mu$ vanish, while their space components reduce to the standard 3-D electric and magnetic fields. In addition, as shown by Eqs.(\ref{1},\ref{2a}) the electric part vanishes if the ideal Ohm's law  holds. In this case we can use $e_\mu =0$ in order to express $b_\mu$ in terms of $ {\vec B}$ and ${\vec v}$ only and magnetic connections, defined  by the   cuts at constant time of the 2D-hypersurfaces generated by  the 4-vectors $b_\mu$ and $u_\mu$ are preserved. \bigskip

In order to search for  generalized  connections when $e_\mu \not = 0$, we  consider the magnetic part of $F_{\mu\nu}$, {   introduce a generalised magnetic 4-vector field   $ {\hat  b}_\mu \equiv b_\mu + d_\mu$  and  define the generalised field tensor} 
\begin{equation}\label{3}
F_{\mu\nu}^b =  {\varepsilon_{\mu\nu\lambda\sigma} {\hat b}_{\lambda}  u_{\sigma}}.
\end{equation} 
Then we look  for the conditions such that $F_{\mu\nu}^b$    satisfies the homogeneous Maxwell's  equations 
\begin{equation}\label{4}
\partial _\mu {\cal F}_{\mu\nu}^b = 0
\end{equation}
where  the dual tensor $ {\cal F}_{\mu\nu}^b$ is defined by 
\begin{equation}\label{4a}
{\cal F}_{\mu\nu}^b \equiv   \varepsilon_{\mu\nu\lambda\sigma} { F}^b_{\lambda\sigma}/2   =  u_{\mu}{\hat b}_{\nu}   - {\hat b}_{\mu} u_{\nu}.
\end{equation}
Following the usual procedure where the homogeneous Maxwell's equations allow us to define  the 4-vector potential  we set 
\begin{equation}\label{4b}	
{F}_{\mu\nu}^b \equiv  \partial _\mu A^b_\nu - \partial _\nu A^b_\mu  =  \partial _\mu A_\nu - \partial _\nu A_\mu + \partial _\mu s_\nu - \partial _\nu s_\mu  ,
\end{equation} 
where  $ A_\mu $ is the  4-vector potential that defines $F_{\mu\nu}$, while $ A^b_\mu $ is a generalized   4-vector potential and $s_\mu \equiv A^b_\mu - A_\mu $ is a ``gauge'' field.
Combining  Eqs.(\ref{2},\ref{3}) and (\ref{4b}) gives
\begin{equation}\label{4c}  {\varepsilon_{\mu\nu\lambda\sigma}  \,  d_{\lambda}  u_{\sigma}} +
{u_{\mu}e_{\nu} - u_{\nu}e_{\mu}}   =   - \partial _\mu s_\nu + \partial _\nu s_\mu .
\end{equation} 
Contracting Eq.(\ref{4c})  with  $u_\mu$ we obtain the compatibility condition \begin{equation}\label{4d}
e_{\mu}   =    u_\nu \partial _\nu s_\mu - u_\nu  \partial _\mu s_\nu   = \partial_\tau s_\mu - u_\nu  \partial _\mu s_\nu ,
\end{equation} 
with $\partial_\tau \equiv u_\mu \partial _\mu$ the convective derivative with respect to the proper time $\tau$, while the remaining components of Eq.(\ref{4c})  determine 
the 4-vector $d_\mu$  which can be obtained from Eq.(\ref{4c})  by contracting it with $u_\alpha \varepsilon_{\alpha\beta\mu\nu}$ and using $d_\mu u_\mu = 0$.
Any choice of the 4-vector field $s_\mu$ compatible with a specified $e_\mu$ in Eq.(\ref{4d}) defines  a generalised  ideal Ohm's law in terms of  modified e.m. fields given 
by the  field tensor    
\begin{equation}\label{4e} {F}_{\mu\nu}^b \equiv {\varepsilon_{\mu\nu\lambda\sigma} {\hat b}_{\lambda}  u_{\sigma}} \equiv 
F_{\mu \nu}  + \partial _\mu s_\nu - \partial _\nu s_\mu,
\end{equation} and generalised conserved connections  defined  by the   cuts at constant time of the 2D-hypersurfaces generated by  the 4-vectors ${\hat b}_\mu$ and $u_\mu$. \\
Note that the choice  where $s_\nu$ is a 4-gradient corresponds  to $e_\mu = 0$  and is
 simply  a standard gauge transformation of the vector potential $ A_\mu $ that does not affect the e.m. fields.

 \section{Relativistic inertial Ohm's law}\label{RIOL}
 
An interesting  choice of the gauge 4-vector $s_\mu$  is 
\begin{equation}\label{5}
s_{\mu}   =   \Pi  u_\mu  
\end{equation} 
with $\Pi$ a scalar field. From  Eq.(\ref{4d}) we obtain 
\begin{equation}\label{5a}
e_{\mu}  = \partial_\tau (\Pi  u_\mu)  + \partial _\mu \Pi= (u_\mu u_\nu +\delta_{\mu\nu})\partial_\nu \Pi  + \Pi \partial_\tau u_\mu \end{equation} 
$$= \partial_\nu [(u_\mu u_\nu +\delta_{\mu\nu}) \Pi] + u_\mu u_\nu (\Pi/n) \partial_\nu n ,
$$
where $ u_\mu u_\nu +\delta_{\mu\nu}$ is the projector perpendicular to $u_\mu$ and the scalar function $n$ is defined by  the continuity equation $\partial_\nu (n u_\nu )= 0$.
Thus we can write 
\begin{equation}\label{5b}
n\, e_{\mu} =   \partial_\nu [  n u_\mu u_\nu  \Pi]  + n  \partial_{\mu} \Pi . \end{equation} 
If we set \,  $\Pi\equiv (P +\epsilon) /(n q),$
where  we interpret $P$ and $\epsilon$ as  the invariant   pressure and  mass-energy density,  $n$ as the invariant numerical density with $m$  and $q$  as the  mass and charge respectively, we obtain 
\begin{equation}\label{5c}
 n  q \, e_{\mu} = n  q \, F_{\mu\nu} u_\nu = \partial_\nu [ \,  u_\mu u_\nu  (P + \epsilon ) \,  ]  + n \,  \partial_{\mu} [(P + \epsilon ) /n] ,\end{equation} 
that gives
\begin{equation}\label{5d}
\ n  q \, e_{\mu}  = \partial _\nu T_{\mu\nu} -   (P/n) \partial_{\mu} n +  n \,  \partial_{\mu} 
(\epsilon  /n) , \end{equation} 
where $T_{\mu\nu} \equiv (P + \epsilon )u_\mu u_\nu  + P \delta_{\mu\nu}  $ can be interpreted as the fluid  energy momentum tensor.
\\ If we further assume that $\epsilon$ and $P$  are  functions of  $n$ only, consistently with the assumption made in the nonrelativistic case in the Introduction,  and use the thermodynamic relationship (see e.g., Eq.(8) of Ref.\onlinecite{D'A} with   all dependences on the entropy density dropped having in effect assumed that the entropy density of the fluid  is uniform and constant)
\begin{equation}\label{TR}
P  = n\partial\epsilon/\partial n - \epsilon, \end{equation} 
we find that the last two terms in Eq.(\ref{5d}) cancel.
Writing  the dissipationless  relativistic single fluid momentum-energy equation in the form
\begin{equation}\label{6}
\partial _\nu T_{\mu\nu}  = n q \,                  F_{\mu\nu}u_\nu                ,\end{equation}
 we see that, if $\epsilon$ and $P$  are  functions of  $n$ only,  the choice of the gauge field $s_{\mu}   =   [(P +\epsilon) /(n q)]  u_\mu$ defines  conserved generalized connections in agreement with Ref.\onlinecite{AsCo}.
 In  the cold $P=0$ limit  $\Pi = m/q$ and  the combination  $A_\mu + s_\mu$ reduces, aside for a multiplication factor,  to the standard (cold) fluid canonical momentum $ m u_\mu +  qA_\mu$. 
In the non relativistic limit for the electron fluid the generalised e.m. fields  that are obtained from the generalised 4-vector potential  $ A_\mu  +m u_\mu/q$ reduce to the electric and magnetic fields 
${\vec E}_e  $ and $ {\vec B}_e$ defined in the Introduction.  
\section{Relativistic   fluid of spherical tops}\label{spin}

An interesting choice of $s_{\mu} $ that  can be related to the motion of relativistic spherical tops \cite{Regge}, in view of the description of a classical fluid of electrons with an internal degree of freedom (such as spin, see e.g., the recent Ref.{\onlinecite{D'Ath}, ch.7,   and references therein), is 
\begin{equation}\label{S}
s_{\mu}   =  C\, \sigma _{\mu\nu}  \sigma _{\nu\lambda} u_\lambda 
\end{equation} 
with $\sigma _{\mu\nu}$ an antisymmetric matrix function and $C$ a scalar constant. We obtain
\begin{equation}\label{Sa}
e_{\mu}  =C\, [ \partial_\tau (\sigma _{\mu\nu}  \sigma _{\nu\lambda} u_\lambda )   - u_\nu  \partial _\mu  (\sigma _{\nu\beta}  \sigma _{\beta\lambda} u_\lambda )].\end{equation} 
If $ \sigma _{\mu\nu}$ is taken to be constant Eq.(\ref{Sa}) can be written as 
\begin{equation}\label{Sb}
e_{\mu}  =C [  (\sigma _{\mu\nu}  \sigma _{\nu\lambda}  \partial_\tau u_\lambda )   -  \partial _\mu  (\Sigma _{\nu}  \Sigma _{\nu})/2], \end{equation} 
with $ \Sigma _{\mu} =  \sigma _{\mu\nu}  u_\nu$.  The gauge field in Eq.(\ref{S}) is compatible with an equation of motion with a modified inertia term  $ (\sigma _{\mu\nu}  \sigma _{\nu\lambda}  \partial_\tau u_\lambda )  $ and a gradient force  term $-  \partial _\mu  (\Sigma _{\nu}  \Sigma _{\nu})/2$ which ensures the consistency of the constraint $u_\mu u_\mu = -1$.

\section{Relativistic  radiation reaction on a cold fluid plasma}\label{RRc}

A different result can be expected in a ``dissipative'' case  setting   
\begin{equation}\label{RR} s_\mu =  C \, \partial_\tau u _{\mu}. \end{equation} 
This choice requires the introduction in the fluid momentum equation  of the second (proper)  time  convective derivative of the fluid 4-velocity  and could be used  to make a comparison  with   the radiation reaction force\cite{LAD}  on a cold relativistic plasma due to emission of (classical) incoherent high frequency radiation  (see also
 Ref.\onlinecite{Mah} for a thermal relativistic plasma). We obtain 

\begin{equation}\label{RRa} e_\mu = C\, [\partial_\tau (\partial_\tau  u_\mu)  - u_\nu \partial_\mu (\partial_\tau u_\nu) ] =  C\, [\partial_\tau \partial_\tau  u_\mu  + (\partial_\mu   u_\nu )(\partial_\tau u_\nu) ] , \end{equation}
which can be rewritten more transparently as 
\begin{equation}\label{RRb}  e_\mu  = C \,(\delta_{\mu\nu} + u_\mu u_\nu) [\partial_\tau \partial_\tau  u_\nu   -    u_\alpha   \partial_\nu \partial_\tau u_\alpha ], \end{equation}
where $(\delta_{\mu\nu} + u_\mu u_\nu)$ is the projector perpendicular to  $u_\mu$ and $(\partial_\mu   u_\nu )(\partial_\tau u_\nu)$. 
\\
While, taking the electron  distribution function to be a  $\delta$ function in momentum space, the term  in Eq.(\ref{RRb}) that involves the second derivative of the 4-velocity with respect to the proper time $\tau$ can be related to  the single particle Lorentz-Abraham-Dirac (LAD) equation \cite{LAD}, the second involves the coordinate derivatives of the acceleration 4-vector $\partial_\tau u_\mu$.  Thus, the introduction in the 4-momentum equation  of a  LAD-force term does not lead to  a generalised Ohm's law compatible with Eq.(\ref{4}) unless the contribution of the second term vanishes.  Conversely, one can use  Eq.(\ref{RRb}) to split the LAD force into  a term  that defines a  generalised ideal Ohm's law and a term  that cannot be included in such a framework. This is important when looking, as is  done in Sec.\ref{hel}, for conservation laws of the plasma dynamics.

\section{Resistive Ohm's law}\label{CL}

A similar splitting of the term that violates the ideal Ohm's law can in principle be found   in the case of a resistive Ohm's law.

We write  the relativistic covariant form   \cite{Ged}  of the resistive term as 
\begin{equation}\label{Re} e_\mu = \eta \, (\delta_{\mu\nu} + u_\mu u_\nu) j_\nu  =  \eta\,  ( j_\mu - \rho u_\mu ),
\end{equation}
where $\eta$ is a scalar resistivity and $j_\mu$ is the current density four vector, $\rho\equiv u_\mu j_\mu$ is the invariant charge density. The projector operator, which is required in order to satisfy the constraint $e_\mu u_\mu = 0$, subtracts from $j_\mu$ the current density arising from  the charge advected   by the fluid 4-velocity $u_\mu$ which is not affected by resistivity. For the sake of simplicity we take $\eta$ to be a constant. \\
Using the inhomogeneous Maxwell's equation $\partial_\nu F_{\mu\nu} = (4\pi/c)\, j_\mu$, we write Eq.(\ref{Re})  as 
\begin{equation}\label{Rea} e_\mu =
(c \eta /4\pi)\, (\partial_\nu F_{\mu\nu} + u_\mu u_\nu \partial_\alpha  F_{\alpha\nu})    \end{equation} 
and compare it with Eq.(\ref{4d}) that requires 
$ e_\mu  =\partial _\tau s_\mu - u_\nu  \partial _\mu s_\nu
$. 
A possible choice for $s_\mu$, in a sense the counterpart of the choice made in Eq.(\ref{RR}) as it involves  integration with respect to the proper time $\tau'$ along the fluid trajectories  instead of differentiation, is to take 
\begin{equation}\label{Reb}  s_\mu = \int ^\tau d\tau' e_\mu (\tau')= (c \eta /4\pi)\, \int ^\tau d\tau'  (\partial_\nu F_{\mu\nu} + u_\mu u_\nu \partial_\alpha  F_{\alpha\nu})'  
  \end{equation} 
  which leaves the  term  $u_\nu  \partial _\mu s_\nu$ unbalanced.  Similarly  to the result of the  preceding section the unbalanced term depends on the coordinate derivatives of $s_\mu$.

\section{Generalised magnetic helicity}\label{hel}

As already  mentioned,  if the generalised Ohm's law can be written in the form $F_{\mu\nu}^b u_\mu =0 $ with $\partial_\mu {\cal F}_{\mu\nu}^b =0$, it is possible to define  in a covariant way generalised magnetic connections between plasma elements.   In this section we consider the generalisation on magnetic helicity that  in the case of the ideal Ohm's law in  Eq.(\ref{int1}) is represented   by the 4-vector
\begin{equation}\label{He} K_\mu = { \cal F}_{\mu\nu} A_\nu.  \end{equation}
The 4-vector $K_\mu$, which  is defined modulo a 4-divergence because of the standard gauge freedom in the choice of the vector potential ($A_\mu \to A_\mu +\partial_\mu \psi$),  satisfies the continuity equation 
\begin{equation}\label{Hea}\partial_\mu  K_\mu =  {\cal F}_{\mu\nu} F_{\mu\nu}/2 = 0 , \end{equation}
where the last equality holds because of the ideal Ohm's law.  In the framework of the above analysis Eqs.(\ref{He},\ref{Hea}) can be generalised by defining 
\begin{equation}\label{Heb} K^b_\mu = { \cal F}^b_{\mu\nu} A^b_\nu,
\end{equation}
where  ${ \cal F}^b_{\mu\nu} $  and $A^b_\nu$ are defined by Eqs.(\ref{4a}, \ref {4b})
and $K^b_\mu$ satisfies  the continuity equation 
\begin{equation}\label{Hec} \partial_\mu  K^b_\mu  = 
\varepsilon_{\mu\nu\alpha\beta}  \, \partial_\mu [(A_\nu + s_\nu) \partial_\alpha( A_\beta + s_\beta) ] = 0. \end{equation}
Referring  for example to the case of the ideal inertial Ohm's law for cold electrons, we see that this generalised continuity equation involves the conservation of the sum of the magnetic helicity defined by  Eq.(\ref{He}}), of a term proportional to the fluid 4-helicity  defined by  $\Omega_{\mu\nu} u_\nu$  where $ \Omega_{\mu\nu} = \varepsilon_{\mu\nu\alpha\beta} \partial_\alpha u_\beta $ is the fluid vorticity and of  two mixed terms proportional to 
$  { \cal F}_{\mu\nu} u_\nu$ and to $ \Omega_{\mu\nu} A_\nu$ respectively.

\section{Conclusions}\label{conc}

The dynamics of relativistic plasmas is a subject of great present interest for both laboratory plasma physics \cite{bulan,nilson}
 and for astrophysical plasmas \cite{HessZen,HoshZen}  and in particular for the conversion of electromagnetic field 
energy into kinetic and thermal energy of the plasma particles and viceversa. In this context the equations of magnetohydrodynamics have been extended (see  Refs.\onlinecite{Li},\onlinecite{Ani})
and used in numerical simulations (see e.g., Ref.\onlinecite{Mign})  so as to include fluid and thermal velocities close to the speed of light  and  the concept of reconnection of magnetic field lines, a fundamental  process in plasmas, has been extended to relativistic regimes \cite{BF}. 
Magnetic reconnection is in fact ubiquitous in magnetised plasmas and  can be viewed  as a process that converts  magnetic energy inside highly inhomogeneous regions  into plasma particle energy and  as a process  that modifies the magnetic topology, more precisely the connections drawn by the magnetic field lines.
These processes are made possible by local effects that are outside the large spatial-scale, long time-interval description of (ideal) MHD theory.
Thus an important point in this relativistic extension of MHD is to provide a frame independent definition of magnetic reconnection. Such a definition  is not obvious both from a theoretical and from an observational point of view, since  the distinction between electric and magnetic fields is frame dependent and the tracing of field lines, which are defined in coordinate space at a given time, is also frame dependent due to the violation of simultaneity in different reference frames of events at different spatial locations.  This point  was  addressed in Refs.\onlinecite{FPEPL},\onlinecite{Arcetri}. 
\\
A second  important point is to find a covariant relativistic extension of the generalised magnetic connections that are known to occur when the ideal Ohm's law is violated by terms that can be accounted for by defining generalised e.m. fields \cite{AsCo}.  Generalised e.m. fields must satisfy: \\
1) an  ideal Ohm's law (see Eq.(\ref{int1})),
\\ 2)  and a set of equations analogous to  the homogeneous Maxwell's equations (see Eq.(\ref{4})).\\
The inclusion of electron inertia terms in the ideal Ohm's equation expressed in terms of the electron fluid velocity is a well studied example \cite{kuv} in the nonrelativistic case.
\\
The   connections  between plasma elements defined by the  generalised electromagnetic fields that satisfy conditions 1) and 2), are conserved by the plasma dynamics and, for the electron inertia case, it was  shown in the literature (see e.g. Refs.\onlinecite{ottav},\onlinecite{Grass},\onlinecite{advan}  and references therein) that they  can have important consequences on the development of magnetic reconnection.  In fact  in this case the generalized magnetic field ${\vec B}_e$ cannot reconnect and thus the  reconnection  of  ${\vec B}$ can only proceed by developing increasingly steeper layers of the electron velocity, and thus of the plasma current density, on scalelength related to the so called electron inertial skin depth. 

In the present article we have examined within a formal framework  whether such extensions can be performed in a covariant relativistic way in terms of a {\it gauge} 4-vector  field 
that adds to  the 4-vector potential field so as to  implement  conditions 1) and 2) automatically.  \\
We have shown that  the extension that  includes the electron inertial term is straightforward  in a cold relativistic plasma and leads to a generalised vector potential that is proportional to the well known fluid canonical momentum. It can easily be extended to include electron thermal effects if the pressure and mass-energy density are assumed to be functions of the electron density only.
\\
We have also exploited the  formal framework  developed in Sec.\ref{ROL}  in order to examine more exotic situations, such as the equation for relativistic spherical  tops  with the aim of looking how to include in this formalism  internal degrees of freedom of the particles in the plasma.
\\
Dissipative terms ranging from a relativistic formulation of resistivity (frequently used in the study of relativistic reconnection, see e.g., Ref.\onlinecite{relsim})  and of the radiation reaction force (of interest for present laser plasma interactions \cite{Tamb}  and suggested as a mechanism for relativistic reconnection in Ref.\onlinecite{Hosh}) have been examined.  It appears that only  part of their contributions can be accounted for by generalised e.m. fields and that the part that cannot be accounted for involves coordinate derivatives and is thus related  to inhomogeneities in the dissipation process between neighbouring plasma elements. 

Finally in Sec.\ref{CL} we have shown that, when  generalised e.m. fields can be defined according to the requirements 1) and 2),  a generalised helicity 4-vector field   can be constructed  that has vanishing 4-divergence i.e.,  that obeys a conservation law expressed by a continuity equation as is the case for the helicity 4-vector field in the context  of 
ideal MHD.


\begin{thebibliography}{99}

\bibitem{Newc} W.A. Newcomb, {\it Ann. Phys.}, {\bf 3},   347 (1958).


\bibitem{Kurch}   A.S. Kingsep, K.V. Chukbar, V.V. Yan$^{prime}$kov, in {\it Reviews of Plasma Physics}, Consultants Bureau,  Vol. 16,  New York, (1990).

\bibitem{EMHD}   S.V. Bulanov, F. {Pegoraro},  A.S. {Sakharov},
 {\it Phys.  Fluids B}, {\bf 4}, 2499 (1992).


\bibitem{hallMHD} M.J. Lighthill, {\it Phil. Trans. Roy. Soc., London}, {\bf 252A}, 397 (1960).


 
 \bibitem{Yan}
 V. V. Yan$^{\prime}$kov,
{\it Zh. Eksp. Teor. Fiz.},  {\bf 107}, 414 (1995).
 
\bibitem{de} E. Cafaro,  D. Grasso, F. Pegoraro,  F. Porcelli, A. Saluzzi, {\it Phys. Rev. Lett.}, {\bf 80},  4430 (1998).

\bibitem{fn1} {We adopt a simplified notation where Greek indices run from $0$ to $3$, only lower indices are used and the index contraction  involves the Minkowski metric  tensor with the minus sign corresponding to its  time-time ($00$) component.}


\bibitem{FPEPL} F. Pegoraro, {\it EPL}, {\bf 99},  35001 (2012).

   

\bibitem{Arcetri} F. Pegoraro, {\it Relativistic Magnetohydrodynamics
and Relativistic Reconnection}, Arcetri 2014 Workshop on Plasma Astrophysics, unpublished. 



\bibitem{fn2} It can be easily shown that when Eq.(\ref{int2}) holds, the rank of the e.m. tensor  $F_{\mu\nu}$ is two.



\bibitem{AsCo} F.A. Asenjo, L. Comisso,  {\it Phys. Rev. Lett.}, {\bf 114},  115003  (2015).



\bibitem{fn3} {If this  4-velocity coincides  with  the 4-velocity of a plasma component,  the $R_\mu$ term  belongs  to the first class  of violations listed in the introduction (and  thus magnetic connections of elements of this plasma component are preserved in time). If this is not the case magnetic topology is  still  conserved, and  e.g., the two dimensional hypersurfaces mentioned in the introduction can be constructed,   but this  does not lead to dynamically conserved magnetic connections between plasma elements as these move with a different 4-velocity.}

\bibitem{Li} A. Lichnerowicz, in {\it Relativistic Hydrodynamics and Magnetohydrodynamics},
(New York: Benjamin) 1967.

\bibitem{Ani} M. Anile, in {\it Relativistic fluids and magneto-fluids}, Cambridge Monographs on
Mathematical Physics, Cambridge 1989.

 \bibitem{D'A} E. D'Avignon, P.J. Morrison, F. Pegoraro ,  {\it Phys. Rev. D}, {\bf 91},  084050 (2015).
 
 \bibitem{Regge}   A.J. Hanson, T. Regge, {\it Ann. Phys.}, {\bf 87}, 498 (1974)
 
  \bibitem{D'Ath}  E. D'Avignon,  in {\it Aspects of Relativistic Hamiltonian Physics},  PhD Dissertation, The University of Texas at Austin, 2015.
 
  \bibitem{LAD} H.A. Lorentz, in {\it The Theory of Electrons}, (Leipzig, New York, 1909);\, 
M. Abraham, in {\it Theorie der Elektrizit\"at}, Vol. II (Teubner, Leipzig, 1905);\,  P.A.M. Dirac, 
 {\it Proc. Roy. Soc. A}, {\bf  167} 148 (1938).
 
   \bibitem{Mah} V.I. Berezhiani, R.D. Hazeltine,  S.M. Mahajan,  {\it Phys. Rev. E}, {\bf 69}, 056406 (2004).
 

  
\bibitem{Ged} M. Gedalin, {\it Phys. Rev. Lett.}, {\bf 76},  3340 (1996).




 \bibitem{bulan} S.V. Bulanov, T.Zh. Esirkepov, D. Habs, F. Pegoraro, T. Tajima,
{\it Eur. Phys. J. D}, {\bf 55}, 483  (2009).

 \bibitem{nilson}  P.M. Nilson,  L. Willingale,  M. C. Kaluza, C. Kamperidis,  S. Minardi,  M.S. Wei, P. Fernandes,  M. Notley,  S. Bandyopadhyay,  M. Sherlock,  R.J. Kingham,  M. Tatarakis,  Z. Najmudin,  W. Rozmus,  R.G. Evans,  M. G. Haines, A.E. Dangor, K. Krushelnick    {\it Phys. Rev. Lett.}, {\bf 97}, 255001(2006). 


 \bibitem{HessZen}  M. {Hesse}, S. {Zenitani}, 
{\it Phys. Plasmas}, {\bf 14}, 112102 (2007).
 
 
 
 \bibitem{HoshZen}  S. {Zenitani}, M. {Hoshino},     
{\it Astroph. J.},
{\bf 670}, 702 (2007).

    \bibitem{Mign} A. Mignone, G. Bodo, {\it Mon. Not. R. Astron. Soc.}, {\bf 368}, 1040 (2006).
   
    
\bibitem{BF} E.G. Blackman, G.B. Field, {\it Phys. Rev. Lett.}, {\bf 72}, 494 (1994). 
    
    
 \bibitem{kuv} 
B.N. Kuvshinov, F. Pegoraro, T.J. Schep,
{\it Phys. Lett. A}, {\bf 191}, 296 (1994),

  \bibitem{ottav}   M. {Ottaviani}, F.{Porcelli},  {\it Phys. Rev. Lett.}, {\bf 71}, 3802 (1993). 


 \bibitem{Grass}   D. {Grasso}, F. {Califano}, F. {Pegoraro}, F. {Porcelli},  {\it Phys. Rev. Lett.}, {\bf 86}, 5051 (2001). 

\bibitem{advan} F. Porcelli, D. Borgogno, F. Califano, D. Grasso, M. Ottaviani, F. Pegoraro,
{\it Plasma Phys.  Contr.  Fus.}, {\bf 44}, {B389} (2002).

\bibitem{relsim} 
S. Zenitani, M. Hesse, A.  Klimas, {\it Astrophys. J. Lett.} , {\bf 716}, L214 (2010).

\bibitem{Tamb}  M. Tamburini, F. Pegoraro, A. Di Piazza, C.H. Keitel, A. Macchi, 
{\it New J.  Phys},  {\bf 12}, 123005(2010).

 \bibitem{Hosh} C.H. Jaroschek, M. Hoshino, {\it Phys. Rev. Lett.}, {\bf 103},  075002 (2009).

\end{thebibliography}
\end{document}